\documentclass[runningheads]{llncs}

\usepackage[font=footnotesize]{caption}
\usepackage[misc]{ifsym}
\usepackage{amsmath, fancyvrb, comment, centernot, tikz, adjustbox, xspace, multirow, cite}
\usepackage{listings,url}
\usepackage{color}
\usepackage{hyperref}
\usepackage[linesnumbered,noend]{algorithm2e}
\usepackage{ulem}

\usetikzlibrary{positioning}
\definecolor{dkgreen}{rgb}{0,0.6,0}
\definecolor{gray}{rgb}{0.5,0.5,0.5}
\definecolor{mauve}{rgb}{0.58,0,0.82}

\newcommand{\blue}[1]{{{\textcolor{blue}{#1}}}}
\newcommand{\green}{}

\newcommand{\proofs}{}
\newcommand{\introfulltext}{}
\newcommand{\fulltext}{}
\newcommand{\arxivLink}{https://arxiv.org/abs/2108.00885}
\newcommand{\subsectionTightenUpper}{\unless\ifdefined\fullText{\vspace{-5mm}}\fi}
\newcommand{\subsectionTightenLower}{\unless\ifdefined\fullText{\vspace{-2mm}}\fi}
\newcommand{\sectionTightenLower}{\unless\ifdefined\fulltext{\vspace{-2mm}}\fi}

\newcommand{\type}{\mathsf{type}}
\newcommand{\len}{\mathsf{len}}
\newcommand{\sts}{\mathsf{STS}}
\newcommand{\traces}{\mathsf{Traces}}
\newcommand{\states}{\mathsf{States}}
\newcommand{\para}[1]{\vspace{-2mm}\paragraph{\textbf{#1.}}}

\usetikzlibrary{
    arrows,shapes,positioning,decorations.markings,
    decorations.pathreplacing,hobby, automata,snakes}
\tikzset{
    looped/.style={
        decoration={markings,mark=at position 0.999 with {\arrow[scale=2]{>}}},
        postaction={decorate},
        >=stealth
    },
    straight/.style={
        decoration={markings,mark=at position 1 with {\arrow[scale=2]{>}}},
        postaction={decorate},
        >=stealth
    },
    loopedSF/.style={
        decoration={
            markings,
            mark=at position 0.999 with {\arrow[scale=2]{>}},
            mark=at position 0.5 with {\arrow[scale=2]{>}}},
        postaction={decorate},
        >=stealth
    },
    straightSF/.style={
        decoration={
            markings,
            mark=at position 0.999 with {\arrow[scale=2]{>}},
            mark=at position 0.5 with {\arrow[scale=2]{>}}},
        postaction={decorate},
        >=stealth
    },
    triangle/.style = {fill=white, draw=black, regular polygon, regular polygon sides=3 },
    node rotated/.style = {rotate=180},
    border rotated/.style = {shape border rotate=180}
}

\newcommand{\Alice}{\texttt{Alice}\xspace}
\newcommand{\Bob}{\texttt{Bob}\xspace}
\newcommand{\Eve}{\texttt{Eve}\xspace}
\newcommand{\Plaintext}{\texttt{Plaintext}\xspace}
\newcommand{\Encrypted}{\texttt{Encrypted}\xspace}

\newcommand{\Process}{\texttt{Process}\xspace}
\newcommand{\Secret}{\texttt{Secret}\xspace}
\newcommand{\Key}{\texttt{Key\textsubscript{AB}}\xspace}
\newcommand{\Message}{\texttt{Message}\xspace}
\newcommand{\Any}{\texttt{*}\xspace}
\newcommand{\True}{\top}
\newcommand{\False}{\bot}

\newcommand{\lt}[1]{\stackrel{#1}{\longrightarrow}}

\newcommand{\always}{\mathbf{G}}

\newcommand{\classify}{\mathsf{classify}}
\newcommand{\verify}{\mathsf{verify}}
\newcommand{\counterexample}{\mathsf{counterexample}}
\newcommand{\facts}{\mathsf{facts}}
\newcommand{\traceConstraint}{\mathsf{traceConstraint}}
\newcommand{\minimizetc}{\mathsf{minimizeTC}}
\newcommand{\redundancyCheck}{\mathsf{removeRedundant}}

\newcommand{\ok}{\mathsf{OK}}
\newcommand{\minCore}{\mathsf{minCore}}
\newcommand{\block}{\mathsf{block}}
\newcommand{\nok}{\mathsf{Violated}}

\newcommand{\Server}{\texttt{Server}\xspace}
\newcommand{\replay}{\mathsf{replay}}
\newcommand{\manInTheMiddle}{\mathsf{manInTheMiddle}}

\newcommand{\Generic}{\mathsf{Generic}}

\begin{document}
\title{Counterexample Classification\thanks{
This work has been supported by the National Science Foundation under NSF SaTC award CNS-1801546.}}

\author{
        Cole Vick\inst{1} \and
        Eunsuk Kang\inst{2} \and
        Stavros Tripakis\inst{1} 
}

\institute{
        Northeastern University \texttt{\{vick.c\},\{stavros\}@northeastern.edu} \and
        Carnegie Mellon University \texttt{eskang@cmu.edu} 
}

\maketitle

\begin{abstract}
    In model checking, when a given model fails to satisfy the desired specification, a typical model checker provides a \emph{counterexample} that illustrates how the violation occurs.
    In general, there exist many diverse counterexamples that exhibit distinct violating behaviors, which the user may wish to examine before deciding how to repair the model.
    Unfortunately, obtaining this information is challenging in existing model checkers since (1) the number of counterexamples may be too large to enumerate one by one, and (2) many of these counterexamples are redundant, in that they describe the same type of violating behavior.
    In this paper, we propose a technique called \emph{counterexample classification}.
    The goal of classification is to partition the space of all counterexamples into a finite set of \emph{counterexample classes}, each of which describes a distinct type of violating behavior for the given specification.
    These classes are then presented as a summary of possible violating behaviors in the system, freeing the user from manually having to inspect or analyze numerous counterexamples to extract the same information. 
    We have implemented a prototype of our technique on top of an existing formal modeling and verification tool, the Alloy Analyzer, and evaluated the effectiveness of the technique on case studies involving the well-known Needham-Schroeder protocol with promising results.
\end{abstract}
        
\section{Introduction}
\label{sec_intro}
\sectionTightenLower
In formal verification, \emph{counterexamples} are an invaluable aid for debugging a system model for possible defects. Typically, a counterexample is constructed by a verification tool as a \emph{trace} (i.e.,\ a sequence of states or events) that demonstrates how the system violates a desired property. The user of the tool would then inspect the counterexample for the underlying cause behind the violation and fix the model accordingly.

In practice, there are a number of challenges that the user may encounter while using counterexamples to debug and repair a model. First, a counterexample may contain details that are irrelevant to the root cause of a violation, requiring considerable effort by the user to manually analyze and extract the violating behavior. Second, the user may wish to investigate multiple different types of counterexamples before deciding how to repair the model; this is, however, a challenging task because (1) the number of counterexamples may be too large to enumerate one by one, and (2) many of these counterexamples may be \emph{redundant} in that they describe the same type  of violating behavior. 


This paper proposes a technique called \emph{counterexample classification} as an approach to overcome these challenges. The key intuition behind this approach is that although a typical model contains a very large (or possibly infinite) set of counterexamples, (1) many of these can be considered ``similar'', in that they share a common, violating behavior and (2) this similarity can be captured as a specific relationship between states that is shared by these traces. Based on this insight, our technique automatically partitions the set of counterexamples into a finite number of \emph{classes}, each of which is associated with a \emph{constraint} that characterizes one particular type of violation. These constraints are then presented to the user, along with representative counterexamples, as distinct descriptions of possible defects in the system, freeing them from manually sorting through numerous counterexamples to extract the same information.
\ifdefined\introfulltext{
        \green{
        For instance, consider a security protocol involving a pair of agents that communicate over a channel, with an attacker that attempts to compromise the secrecy of exchanged information by carrying out various attacks.
        Although a model of the protocol may admit a large number of counterexample traces (each corresponding to a possible attack), suppose that each attack on this protocol can be classified as an instance of (1) a \emph{man-in-the-middle} attack where the attacker places itself between the two agents or (2) a \emph{replay} attack where the attacker exploits a previously sent message. Given this model, our technique would automatically generate and present these two classes to the user together with representative counterexamples from each class.}
}\fi

\ifdefined\introfulltext{
        \green{
        A key idea behind our approach is the use of user-defined summary \emph{predicates} for classifying counterexamples.
        In certain domains, the user may have a priori knowledge about common types of defects that can be encoded as generic constraints over system primitives.
        For example, there are well-understood categories of security attacks (e.g.,\ man-in-the-middle and replay attacks) that can be expressed over  concepts such as keys, messages, and agents.
        Our approach allows the user to control and fine-tune the result of classification.
        In addition, once defined, these predicates may be reused across multiple models within the same domain, as we demonstrate with the verification of security protocols in this paper.}
}\fi

We have built a prototype implementation of our classification technique on top of an existing formal modeling and verification tool, the Alloy Analyzer~\cite{jackson_alloy_2002}. Our tool accepts a formal model, a specification (that the model currently violates), and a set of \emph{predicates} that describe relationships between states in the model. From these, the tool produces (if one exists) a set of classes that accounts for all of the violating behavior in the model. As a case study, we have successfully applied our technique to two variants of the Needham-Schroeder protocol~\cite{needham_using_1978}, and were able to classify hundreds of thousands of counterexamples into only a handful of classes that represent known attacks to the protocol.

Our main contributions may be summarized as follows: a formal definition of the Counterexample Classification Problem (Section 3),  a solution to the Counterexample Classification Problem (Section 4), and a case study on a well-established distributed protocol, Needham-Schroeder (Section 5), that demonstrates the efficacy of our solution.   

\subsectionTightenUpper
\subsection{Running Example}
\label{sec_running_example}
\subsectionTightenLower

To motivate our technique, we introduce the following example. \Alice and \Bob are sending \Message{s} to each other. \Eve is able to view these messages as they are being sent. The content of a message can be either \Plaintext or \Encrypted. \Eve is always able to read \Plaintext messages, but needs \Key, Alice and Bob's shared key, to read \Encrypted messages. \Eve acquires \Key by seeing an \Encrypted message, modelling \Eve ``breaking'' the encryption of what should be a one-time key. A \Message may be flagged as \Secret, meaning that its content should not be read by \Eve.

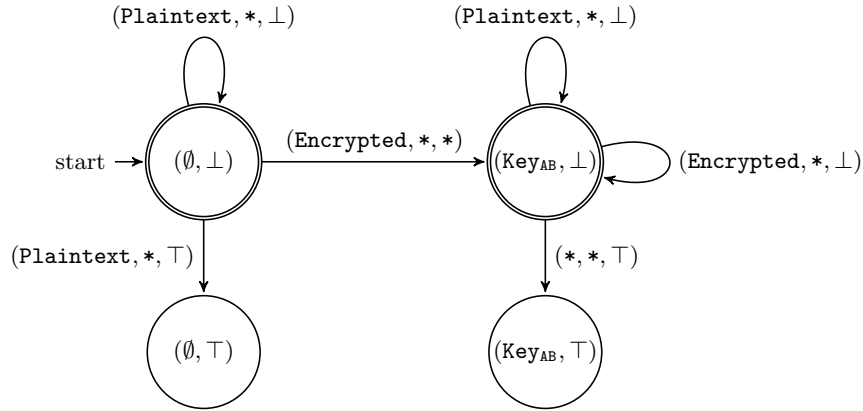
\begin{figure}
  \centering
  \begin{adjustbox}{max totalsize={1.0\textwidth}{.3\textheight}}
    \begin{tikzpicture}[->, >=stealth', shorten >=1pt, auto, semithick]
      
      \tikzstyle{every state} = [minimum size=1.5cm, inner sep=0pt, align=center]
      \tikzstyle{every node} = [font=\small]
      
      \node[state, initial, accepting]  (init) 
           {$(\emptyset, \False)$};
           \node[state, accepting]  (learnKey) 
                [right=3cm of init] 
                {$(\Key, \False)$};
                \node[state]  (learnKeyViolation)
                     [below=1cm of learnKey]
                     {$(\Key, \True)$};
                     \node[state]  (noKeyViolation) 
                          [below=1cm of init]
                          {$(\emptyset, \True)$};
                          
                          \path (init) 
                          edge [loop above] node[pos=0.5, above]
                          {$(\Plaintext, \Any, \False)$}
                          (init)
                          edge [above]  node[pos=0.5, above]
                          {$(\Encrypted, \Any, \Any)$}
                          (learnKey)
                          edge [below]  node[pos=0.5, left]
                          {$(\Plaintext, \Any, \True)$}
                          (noKeyViolation)
                          (learnKey)   
                          edge [loop above] node[pos=0.5, above]
                          {$(\Plaintext, \Any, \False)$}
                          (learnKey)
                          edge [loop right] node[pos=0.5, right]
                          {$(\Encrypted, \Any, \False)$}
                          (learnKey)
                          edge [below] node[pos=0.5, right]
                          {$(\Any, \Any, \True)$}
                          (learnKeyViolation);              
    \end{tikzpicture}
  \end{adjustbox}

  \caption{Transition system of the running example.}
  \label{fig:publicChan}
\end{figure}

We model this example as a transition system, shown in Figure~\ref{fig:publicChan}. The transition system has four states, represented by two state variables, $EveKey$ of type $Key = \{\emptyset, \Key\}$, and $EveSeenSecret$ of type Boolean ($\True$ for true and $\False$ for false).
\ifdefined\introfulltext{
\green{$EveKey=\Key$ means that \Eve has learned the key shared by \Alice and \Bob, while $EveKey=\emptyset$ means that \Eve does not know the key. $EveSeenSecret=\True$ means that \Eve has read a secret message.}
}\fi
The initial state is $(\emptyset,\False)$ meaning that \Eve does not know the key and has not read any secret.

Transitions between states are labeled by \Message{s}. A \Message is a tuple of the form $(type,sender,secret)$, where
$type\in\{\Encrypted,\Plaintext\}$ denotes whether the message is encrypted or not (if encrypted, a message is encrypted by $\Key$), 
$sender\in \{\Alice,\Bob\}$ denotes the sender of the message, 
and $secret$ is a Boolean denoting whether the message is secret or not.
For example, the transition $(\emptyset,\False) \lt{(\Plaintext,\Alice,\True)} (\emptyset,\True)$ means that \Alice sends a \Plaintext (unencrypted) \Secret message.\footnote{The traces in this section have labels, ie. \Message{s}, on their transitions. We do this to make it clear how messages are sent and how different messages affect the state. Our formal definition will not include labels as they may be encoded directly into the state.
}
\Any indicates that the corresponding field can take any value within its type, i.e., there are multiple such transitions, one for each possible value.

We would like this system to satisfy the property that \Eve never reads a \Message that is flagged as \Secret. This can be expressed as the temporal logic (LTL) formula
$$\Phi = \always(EveSeenSecret = \False)$$ which states that $EveSeenSecret=\False$ holds at every reachable state of the system, i.e., it is an {\em invariant}.
As we can see, this is not the case for the model in Figure~\ref{fig:publicChan}.
\ifdefined\introfulltext{\green{The two top states satisfy the property (these are the {\em good} states), whereas the two states at the bottom of the figure do not (these are the {\em bad} or {\em error} states).}}
\else
The top two double bordered states are the {\em good} states.
\fi

Note that this system has infinitely many {\em counterexample traces}, as self-loop transitions can be taken arbitrarily many times.
\ifdefined\introfulltext{\green{Even when a system has a finite number of violating traces, presenting all of them to the user is not a good idea, as there are typically far too many to analyze. Keeping that in mind, some of the questions examined in this paper are the following: {\em How many of the violating traces should be presented to the user as counterexamples?} {\em Are some of these counterexamples {\em similar} in some sense?} {\em Can they be {\em classified} into some type of similarity {\em classes} so that only those classes are presented to the user?}}}
\fi

Take for instance the counterexample traces listed below:
\begin{eqnarray*}
\rho_1^1 & = & (\emptyset,\False) \lt{(\Plaintext,\Alice,\True)} (\emptyset,\True) \\
\rho_1^2 & = & (\emptyset,\False) \lt{(\Plaintext,\Alice,\False)} (\emptyset,\False) \lt{(\Plaintext,\Alice,\True)} (\emptyset,\True) \\
\rho_1^3 & = & (\emptyset,\False) \lt{(\Plaintext,\Bob,\False)} (\emptyset,\False) \lt{(\Plaintext,\Bob,\True)} (\emptyset,\True) \\
\rho_1^4 & = & (\emptyset,\False) \lt{(\Plaintext,\Alice,\False)} (\emptyset,\False) \lt{(\Plaintext,\Bob,\True)} (\emptyset,\True) 
\end{eqnarray*}
In $\rho_1^1$, \Alice sends a \Plaintext \Secret message. \Eve is be able to read it, as it is unencrypted, which leads to a violation of the property. 
In $\rho_1^2$, \Alice first sends a \Plaintext but non-secret message and then sends a \Plaintext \Secret message.
In $\rho_1^3$,  \Bob first sends a \Plaintext but non-secret message and then he sends a \Plaintext \Secret message.
In $\rho_1^4$, \Alice sends a \Plaintext but non-secret message and then \Bob sends a \Plaintext \Secret message. 

These violating traces share important behavior: the fact that either \Alice or \Bob sends a \Plaintext \Secret message. Noticing this, we would like to group these traces together in the same {\em counterexample class}.
\ifdefined\introfulltext{\green{Note that this class contains not only the above four counterexamples, but an infinite number of distinct counterexamples where \Alice or \Bob sends a \Plaintext \Secret message. A potential succinct description of the class stated in words may be: \Eve receives a \Plaintext \Secret message sent by \Alice or \Bob.}}\fi

Now consider the counterexample traces listed below:

\ifdefined\introfulltext{
\begin{eqnarray*}
  \rho_2^1 & = & (\emptyset,\False) \lt{(\Encrypted,\Alice,\False)} (\Key,\False) \lt{(\Encrypted,\Alice,\True)} (\Key,\True) \\
  \rho_2^2 & = & (\emptyset,\False) \lt{(\Encrypted,\Bob,\False)} (\Key,\False) \lt{(\Encrypted,\Alice,\True)} (\Key,\True) \\
  \rho_2^3 & = & (\emptyset,\False) \lt{(\Encrypted,\Alice,\True)} (\Key,\False) \lt{(\Encrypted,\Alice,\True)} (\Key,\True) \\
  \rho_2^4 & = & (\emptyset,\False) \lt{(\Encrypted,\Bob,\True)} (\Key,\False) \lt{(\Encrypted,\Alice,\True)} (\Key,\True) \\
  \rho_2^5 & = & (\emptyset,\False) \lt{(\Encrypted,\Alice,\False)} (\Key,\False) \lt{(\Encrypted,\Bob,\True)} (\Key,\True) \\
  \rho_2^6 & = & (\emptyset,\False) \lt{(\Encrypted,\Bob,\False)} (\Key,\False) \lt{(\Encrypted,\Bob,\True)} (\Key,\True) \\
  \rho_2^7 & = & (\emptyset,\False) \lt{(\Encrypted,\Alice,\True)} (\Key,\False) \lt{(\Encrypted,\Bob,\True)} (\Key,\True) \\
  \rho_2^8 & = & (\emptyset,\False) \lt{(\Encrypted,\Bob,\True)} (\Key,\False) \lt{(\Encrypted,\Bob,\True)} (\Key,\True) 
\end{eqnarray*}
}\fi
\unless\ifdefined\introfulltext{
\vspace{-4mm}
\begin{eqnarray*}
  \rho_2^1 & = & (\emptyset,\False) \lt{(\Encrypted,\Alice,\False)} (\Key,\False) \lt{(\Encrypted,\Bob,\True)} (\Key,\True) \\
  \rho_2^2 & = & (\emptyset,\False) \lt{(\Encrypted,\Bob,\False)} (\Key,\False) \lt{(\Encrypted,\Alice,\True)} (\Key,\True) \\
  \rho_2^3 & = & (\emptyset,\False) \lt{(\Encrypted,\Alice,\True)} (\Key,\False) \lt{(\Encrypted,\Alice,\True)} (\Key,\True) \\
  \rho_2^4 & = & (\emptyset,\False) \lt{(\Encrypted,\Bob,\True)} (\Key,\False) \lt{(\Encrypted,\Alice,\True)} (\Key,\True) 
\end{eqnarray*}
}\fi

These traces exhibit a different way in which the property can be violated than the traces shown previously. Now, the violation happens when \Alice or \Bob send an \Encrypted \Secret message after an \Encrypted message has already been sent, i.e. after \Eve has broken the encryption. A description of this new class would be: \Eve receives an \Encrypted message before receiving an \Encrypted \Secret message.

The method and tool presented in this paper generate such counterexample classes automatically. Our tool does not output class descriptions in English but represents classes syntactically as \emph{trace constraints}. A trace constraint is evaluated over a given trace $\rho$. If $\rho$ satisfies the trace constraint then we say that $\rho$ falls into the class that the trace constraint represents. The trace constraints that represent the two classes discussed above are: 

\vspace{-4mm}
\begin{eqnarray*}\label{tc-running-ex}
  TC_{Plaintext} \label{tcplaintext}
  [\rho] & \equiv & \exists i \in [0 .. len(\rho)] : \rho.type@i = \Plaintext \land \rho.secret@i = \True \\
  TC_{Encrypted} \nonumber
  [\rho] & \equiv & \exists i, j \in [0 .. len(\rho)] : i < j \land \rho.EveKey@i = \Key \land \\ \label{tcencrypted}
					&	& \qquad\qquad\qquad \rho.type@j = \Encrypted \land \rho.secret@j = \True
\end{eqnarray*}
where $len(\rho)$ denotes the length of trace $\rho$ and
the variables $i$ and $j$ represent indices to particular positions of states and transitions in $\rho$. 
The initial state is indexed at position $s_0$ and the first transition is indexed at position $l_0$ and leads to state $s_1$ thus following the general pattern: $s_0\lt{l_0}s_1\lt{l_1}s_2\cdots$.

\ifdefined\introfulltext{\green{We remark that the fact that the two classes above have a 1:1 correspondence with the two error states in the automaton of Figure~\ref{fig:publicChan} is coincidental and not a feature of our technique.
Later, we will present examples of classifications that break this correspondence both for this small example and our larger case study.}}\fi

\sectionTightenLower

\section{Background}
\sectionTightenLower
\begin{definition}[Symbolic transition system]
A {\em symbolic transition system} is a tuple $(X,I,T)$ where:
\begin{itemize}
\item $X$ is a finite set of {\em typed state variables}. Each variable $x\in X$ has a type, denoted $\type(x)$. A type is a set of {\em values}. 
\item The {\em initial state predicate} $I$ is a predicate (i.e., Boolean expression) over $X$.
\item The {\em transition relation predicate} $T$ is a predicate over $X\cup X'$, where $X'$ denotes the set of {\em primed} (next state) variables obtained
from $X$. For example, if $X = \{x,y,z\}$ then $X' = \{x',y',z'\}$. Implicitly, every primed variable has the same
type as the original variable: $\forall x \in X : \type(x')=\type(x)$.
\end{itemize}
\end{definition}

We let $U$ denote the universe of all values.
A {\em state} $s$ over a set of state variables $X$ is an assignment of a value (of the appropriate type) to each variable in $X$, i.e., $s$ is a (total) function $s : X \to U$, such that $\forall x \in X : s(x) \in \type(x)$.
A state $s$ satisfies a predicate $I$ over $X$, denoted $s\models I$, if when we replace all variables in $I$ by their values as defined by $s$, $I$ evaluates to true. For example, suppose $X=\{x,y,z\}$ where $x$ and $y$ are integer variables, and $z$ is a Boolean variable. Let $I$ be the predicate $x<y\land z$. Consider two states, $s_1 = (x=3,y=4,z=\True)$ and
$s_2 = (x=3,y=1,z=\True)$.
Then, $s_1\models I$ but $s_2\not\models I$.

Similarly, a pair of states $(s,s')$ satisfies a predicate $T$ over $X\cup X'$ if when we replace all variables from $X$ in $T$ by their values as defined by $s$, and all variables from $X'$ in $T$ by their values as defined by $s'$, $T$ evaluates to true. For example, suppose $X=\{x\}$ where $x$ is an integer variable. Let $T$ be the predicate $x'=x+1$. Consider three states, $s_0 = (x = 0)$, $s_1 = (x = 1)$, and $s_2 = (x = 2)$. Then $(s_0,s_1)\models T$ and $(s_1,s_2)\models T$, but $(s_0,s_2)\not\models T$.

\begin{definition}[Transition system defined from a symbolic transition system]
A symbolic transition system $(X,I,T)$ defines a {\em transition system} $(S,S_0,R)$, where:
\begin{itemize}
\item The set of {\em states} $S$ is the set of all assignments over $X$.
\item The set of {\em initial states} $S_0$ is the set: $S_0 = \{s \in S \mid s \models I\}$.
\item The {\em transition relation} $R$ is the set: $R = \{(s,s')\in S\times S \mid (s,s')\models T\}$.
\end{itemize}
\end{definition}

That is, the set of initial states is the set of all states satisfying $I$, and the transition relation $R$ is the set of all pairs of states satisfying $T$. A pair $(s,s')\in R$ is also called a {\em transition}, and is sometimes denoted $s\to s'$.

\begin{definition}[Trace]
\label{def_trace}
A {\em trace} $\rho$ over a set of state variables $X$ is a finite sequence of states over $X$:
$\rho = s_0, ..., s_k$. 
The {\em length} of $\rho$ is $k$, and is denoted by $\len(\rho)$; note that $k$ may equal $0$, in which case the trace is empty. 
The set of states of $\rho$ is $\{s_0,...,s_k\}$ and is denoted $\states(\rho)$.
\end{definition}

\begin{definition}[Property]
A {\em property} $\Phi$ over a set of state variables $X$ is a set of traces over $X$.
\end{definition}

\begin{definition}[Traces for an STS]
Let $\sts = (X,I,T)$ be a symbolic transition system and let $(S,S_0,R)$ be the transition system of $\sts$.
The {\em set of traces generated by $\sts$}, denoted $\traces(\sts)$, is the set of all traces $\rho = s_0, s_1, ..., s_k$ over $X$ such that: 
\begin{itemize}
\item $s_0\in S_0$. That is, $\rho$ starts at an initial state of $\sts$.
\item $\forall i\in\{0,...,k-1\} : (s_i,s_{i+1})\in R$. That is, every pair of successive states in $\rho$ is linked by a transition in $\sts$.
\end{itemize}
\end{definition}

\begin{definition}[Property satisfaction and counterexamples]
Let $\sts = (X,I,T)$ be a symbolic transition system and let $\Phi$ be a property over $X$.
We say that $\sts$ {\em satisfies} $\Phi$, written $\sts\models\Phi$, iff $\traces(\sts) \subseteq \Phi$.
If $\sts\not\models\Phi$, then a {\em counterexample} is any trace $\rho\in\traces(\sts)\setminus\Phi$, i.e.,
any trace of $\: \sts$ which violates (does not belong in) $\Phi$.
\end{definition}

\section{Counterexample Classification}
\sectionTightenLower
\subsection{Classes and Classifications}\label{sec_classes}
\subsectionTightenLower

Consider a set of traces $P$.  A \emph{class of $P$} is any non-empty subset of $P$.  A \emph{classification of $P$} is a partition of $P$ into (not necessarily disjoint) classes.

\begin{definition}[Classification]\label{def:classification}
Consider a set of traces $P$. 
A \emph{classification of $P$} is a finite set $C$ of classes of $P$ such that $\bigcup_{c \in C} c = P$. 
\end{definition}

Given a set of counterexample traces $P$, and a classification $C$ of $P$, 
a \emph{canonical counterexample} is a counterexample trace that belongs in exactly one class of $C$. A canonical counterexample thus represents the violating behavior of a particular class as it only appears in that particular class. 

\begin{definition}[Canonical Counterexample]\label{canonicalCounterexample}
  Given a set of counterexamples traces $P$ and a classification $C$ of $P$, a \emph{canonical counterexample} $\rho$ is any counterexample in $P$ such that:
    $\forall c_1,c_2 \in C : (\rho \in c_1 \land \rho \in c_2) \to c_1=c_2$.
We denote by $c(\rho)$ the unique class in $C$ that $\rho$ belongs to.
\end{definition}

A classification is {\em redundant} if it contains classes that have no canonical counterexample:

\begin{definition}[Redundant Classification]\label{def_redundancy}
  A classification $C$ of a set of counterexamples $P$ is \emph{redundant} if there exists a class $c\in C$ such that $c$ does not contain a canonical counterexample.
\end{definition}

\begin{example}
Suppose $P=\{\rho_1,\rho_2,\rho_3,\rho_4,\rho_5\}$ and $C=\{c_1,c_2,c_3\}$ with
$c_1 = \{\rho_1, \rho_2, \rho_3\}, c_2 = \{\rho_3, \rho_4, \rho_5\}, c_3 = \{\rho_1, \rho_4\}$. 
Note that $C$ is a valid classification of $P$ as $c_1\cup c_2\cup c_3 = P$. 
$C$ is a redundant classification, because although $c_1$ has a canonical counterexample $\rho_2$, and $c_2$ has canonical counterexample $\rho_4$, $c_3$ has no canonical counterexample. 
\end{example}

Often, we would like for a classification to guarantee that each class has a canonical counterexample, i.e., to be \emph{non-redundant}. In general, we can transform every redundant classification into a non-redundant classification. First, we state the following two 
\ifdefined\proofs
lemmas:
\else
lemmas\footnote{Proofs for the following Lemmas and Theorems have been removed due to page restrictions. The full paper, with proofs, is available here~\url{\arxivLink}.}:
\fi

\begin{lemma}
\label{lem_redundant}
A classification $C$ of a set of counterexamples $P$ is redundant iff there exist distinct classes $c,c_1,...,c_n\in C$ such that $c \subseteq \bigcup_{i=1,...,n} c_i$.
\end{lemma}

\ifdefined\proofs
\begin{proof}
($\Leftarrow$) 
Suppose there exist distinct classes $c,c_1,...,c_n\in C$ such that $c \subseteq \bigcup_{i=1,...,n} c_i$.
We claim that $c$ has no canonical counterexample. Indeed, take an arbitrary $\rho\in c$. 
Since $c \subseteq \bigcup_{i=1,...,n} c_i$, there must be some $c_i$ such that $\rho\in c_i$.
Moreover, $c$ and $c_i$ are distinct. Therefore, $\rho$ cannot be canonical. 
Since $\rho$ was chosen arbitrarily, there is no canonical counterexample in $c$, which means that $C$ is redundant.

($\Rightarrow$)
Suppose $C$ is redundant. Then there exists $c\in C$ such that $c$ has no canonical counterexample.
By definition, $c$ is non-empty, so pick a $\rho \in c$. By assumption, $\rho$ is not canonical.
Therefore, there exists another class $c'\in C$, distinct from $c$, such that $\rho\in c'$.
Let us denote $c'$ by $c_\rho$, for any arbitrary $\rho$ in $c$. Then $c \subseteq \bigcup_{\rho\in c} c_\rho$.
Moreover, the number of classes in $C$ is finite, so even if $c$ is an infinite set, the set of classes $\{c_\rho\}_{\rho \in c}$ is finite. Call that set $\{c_1,...,c_n\}$. 
Then $\bigcup_{\rho\in c} c_\rho = \bigcup_{i=1,...,n} c_i$, and thus $c\subseteq \bigcup_{i=1,...,n} c_i$.
\qed
\end{proof}

\fi

\begin{lemma}
\label{lem_redundant2}
Let $C = \{c_1,...,c_n\}$ be a classification of a set of counterexamples $P$. $C$ is redundant iff there exists $i\in\{1,...,n\}$ such that $c_i \subseteq \bigcup_{j\ne i} c_j$.
\end{lemma}

\ifdefined\proofs{

\begin{proof}
($\Leftarrow$) 
Follows directly from Lemma~\ref{lem_redundant}.

($\Rightarrow$)
Suppose $C$ is redundant. Then by Lemma~\ref{lem_redundant}, there exist distinct classes $c_i,c_{k_1},...c_{k_m}\in C$ such that $c_i\subseteq \bigcup_{j=1,...,m} c_{k_j}$.
But $\bigcup_{j=1,...,m} c_{k_j} \subseteq \bigcup_{j\ne i} c_j$, therefore,
$c_i\subseteq  \bigcup_{j\ne i} c_j$.
\qed
\end{proof}

}\fi

Based on Lemma~\ref{lem_redundant2}, we can construct an algorithm to transform any classification into a non-redundant classification.
\ifdefined\proofs{
Indeed, let $C$ be a classification, where $C = \{c_1,...,c_n\}$.
First, we iterate over $i$ and check whether there exists an $i$ such that $c_i \subseteq \bigcup_{j\ne i} c_j$.
If no such $i$ exists, then, by Lemma~\ref{lem_redundant2}, $C$ is not redundant and we are done. 
If such an $i$ does exist, then we remove $c_i$ from $C$, to obtain the new classification $C_1 = C \setminus \{c_i\}$.
Note that by removing $c_i$ we do not run the risk of not covering the entire set of counterexamples $P$, since $c_i$ is contained in the union of the remaining classes. 
We continue in this way, removing any class that is covered by the union of all the other classes, until no such class exists, resulting in a non-redundant classification.
Note that the procedure is efficient because in the worst case, we perform no more than $n$ checks of the form
$c_i \subseteq \bigcup_{j\ne i} c_j$, where $n$ is the number of classes in the original classification $C$.
}\fi

\subsectionTightenUpper
\subsection{The Counterexample Classification Problem}\label{subsec_ccp}
\subsectionTightenLower

In Section~\ref{sec_classes}, we defined the concepts of classes and classifications {\em semantically}.
But in order to define the counterexample classification problem that we solve in this paper, we need a {\em syntactic} representation of classes. We define such a representation in this section, by means of {\em trace constraints}.
%
%
A trace constraint is a special kind of predicate that evaluates over traces. A trace constraint is similar to predicates such as the $I$ (initial state) predicate of a symbolic transition system, with two key differences: (1) a trace constraint is only conjunctive, and (2) a trace constraint can refer to state variables at certain positions in the trace and impose logical conditions over those positions. For example, if $X=\{x,y\}$ is the set of state variables, then here are some examples of trace constraints:
\begin{itemize}
\item $TC1[\rho] \equiv \exists i \in [0 .. len(\rho)] : x@i = y@i$: this trace constraint says that there is a position $i$ in the trace such that the value of $x$ at that position is the same as the value of $y$. 
\item $TC2[\rho] \equiv \exists i, j \in [0 .. len(\rho)] : i<j \land x@i > x@j$: this says that there are two positions $i$ and $j$ in the trace such that $i$ is earlier than $j$ and the value of $x$ decreases from $i$ to $j$. 
\end{itemize}

We call formulas such as $x@i = y@i$ or $x@i > x@j$, which operate on indexed state variables, {\em atomic facts}.
We call formulas such as $i<j$, which operate on position variables, {\em atomic position facts}.
Then, a trace constraint is a conjunction of atomic facts and atomic position facts, together with an existential quantification of all position variables within the range of the length of the trace.


Atomic facts and atomic position facts are defined over a set of {\em user-defined predicates}. Some predicates will be standard, such as \emph{equality} ($=$) for integers and \emph{less-than} ($<$) for positions, while other predicates may be domain-specific. In addition to variables, we allow predicates to refer to constants. For example, $i \le 10$ says that the position $i$ must be at most $10$, and $x@2 = 13$ says that the value of $x$ at position $2$ must be $13$. 

For example, recall the \Message type from the running example. The user might want to define a predicate that checks whether two messages have the same sender. Then, the user can define the predicate $SendersEqual$ which is parameterized over two variables of type \Message and defined as:
$$SendersEqual[m_1, m_2] \equiv m_1.sender = m_2.sender$$ 
This predicate may be then instantiated as: 
$$SendersEqual[message@1, message@5]$$ 
This checks whether the \Message at position 1 has the same sender as the \Message at position 5.

\begin{definition}[Trace Constraint]\label{def:traceConstraints}
A trace constraint over a set of state variables $X$ and a set $V$ of 
user-defined predicates is a formula of the form
$$
 TC[\rho] \equiv \exists i_1,...,i_k \in [0 .. len(\rho)] : \xi_0\land\xi_1\land\cdots\land\xi_n
$$
where:
\begin{itemize}
\item $i_1,...,i_k$ are non-negative integer variables denoting positions in the trace $t$. 
We allow $k$ to be $0$, in which case the trace constraint has no position variables. 
\item Each $\xi_j$, for $j=0,...,n$, is either an atomic fact over state variables $X$ and position variables $i_1,...,i_k$ or an atomic position fact over position variables $i_1,...,i_k$ using predicates in $V$. 

\end{itemize}
\end{definition}

Given a trace constraint $w$, and a trace $\rho$, we can evaluate $w$ on $\rho$ in the expected way.
For example, the trace $(x = 0) \longrightarrow (x = 0)$ over state variable $x$, satisfies the trace constraint $TC_1[\rho] \equiv \exists i_0, i_1 \in [0 .. len(\rho)] : i_0 < i_1 \land x@0 = x@1$ but does not satisfy the trace constraint $TC_2[\rho] \equiv \exists i_0, i_1 \in [0 .. len(\rho)] : i_0 < i_1 \land x@0 > x@1$. 
We write $\rho \models w$ if trace $\rho$ satisfies trace constraint $w$. We also say that $w$ \emph{characterizes} $\rho$ when $\rho \models w$.
We denote by $c(w)$ the set of all traces satisfying constraint $w$.

Let $W$ be a set of trace constraints. Then, let $C(W) = \{ c(w) \mid w \in W \}$; i.e.,  $C(W)$ is the set of all sets of traces that are characterized by some trace constraint in $W$. 

Consider a symbolic transition system $\sts$ and a property $\Phi$ that is violated by $\sts$, i.e., $\sts\not\models\Phi$.
The problem that we are concerned with in this paper is to find a classification of all traces of $\sts$ that violate $\Phi$, such that this classification is represented by a set of trace constraints defined over $V$. We call this problem the {\em counterexample classification problem} (CCP):
\begin{definition}[Counterexample Classification Problem]\label{counterexampleClassificationProblem}
  Given symbolic transition system $\sts=(X,I,T)$, property $\Phi$ such that $\sts \not\models \Phi$, and user-defined predicates $V$, find, if there exists, a set of trace constraints $W$ such that:
  (1) each $w \in W$ is a trace constraint over $X$ and $V$; and
  (2) $C(W)$ is a classification of $P$, where $P$ is the set of all traces of $\sts$ that violate $\Phi$.
\end{definition}

\begin{lemma}
Let $W$ be a solution to the CCP. Then, every trace constraint $w \in W$ is a sufficient condition for a violation, i.e., $\forall w \in W : c(w) \cap \Phi = \emptyset$. 
\end{lemma}

\ifdefined\proofs{
  \begin{proof}
    Recall that a classification $C$ of a set of $P = \traces(\sts) \setminus \Phi$ must satisfy $\bigcup_{c \in C} c = P$ (Definition ~\ref{def:classification}).
    Assume $W$ is a solution to the CCP and there exists a $w \in W$ such that $c(w) \cap \Phi \ne \emptyset$. 
    Thus, $w$ accounts for some trace that is not in $P$.
    We've reached a contradiction because if this $w$ were in $W$ then $\bigcup_{w \in W} c(w) \ne P$. 
\qed
\end{proof}

}\fi

\unless\ifdefined\fulltext{
  Note that while a semantic classification always exists, a {\em syntactic} classification, i.e., a solution to CCP in the form of $W$, may not always exist. Whether or not one exists depends on the set of user-defined predicates $V$. Additionally, for a given $V$, CCP may admit more than one valid solution. These topics are treated in further detail in the full paper \cite{arxiv}.
}\fi

\ifdefined\fulltext{
\subsection{Solvability}
\label{subsec-solvability}

\green{
The CCP is formulated as to find a set of trace constraints $W$ {\em if one exists} (Definition~\ref{counterexampleClassificationProblem}). Indeed, while a semantic classification always exists (e.g., a trivial one is the one containing just one class, the set of all counterexamples $P$), a {\em syntactic classification} in the form of $W$ might not always exist. Whether or not one exists depends on the set of user-defined predicates $V$.}

\begin{lemma}
\label{lem_equality}
If the set of counterexample traces $P$ is finite, and $V$ includes equality $=$, then CCP always has a solution.
\end{lemma}

\ifdefined\proofs{
  \begin{proof}\label{proof:solvability}
Recall that a trace is a finite sequence of states (Definition~\ref{def_trace}). 
Then, a trace $s_0, s_1,...,s_k$ can be characterized by a conjunction of $k+1$ formulas, 
$\phi_0\land\phi_1\land\cdots\land\phi_k$, where each $\phi_i$ is itself a conjunction of atomic facts capturing state $s_i$.
Specifically, let $X=\{x_1,...,x_n\}$ be the set of state variables. Then, $\phi_i$ is of the form
$x_1@i = v_1 \land x_2@i = v_2 \land \cdots \land x_n@i = v_n$, where $v_j$ is the value of state variable $x_j$ at state $s_i$.
Notice that each $\phi_i$ has no position variables (Definition~\ref{def:traceConstraints}) because $i$ is instantiated as a constant ranging from $0$ to $k$. 
Indeed, in a fact such as $x_1@i = v_1$, $i$ is the $i$-th position in the trace. 
It follows that $\phi_0\land\phi_1\land\cdots\land\phi_k$ is a trace constraint (without position variables).
Therefore, a single trace can be characterized by a single trace constraint, and thus, such a trace constraint can also represent a class with a single trace in it.
Therefore, if the set of counterexample traces is finite, we can have a classification represented by a finite number of trace constraints, one per counterexample trace. 
\qed
\end{proof}

}\fi


\green{Lemma~\ref{lem_equality} shows that in the presence of equality $=$, and provided that the set of counterexamples is finite, CCP always has a solution.  But in the absence of $=$, CCP may not have a solution.}

\green{For example, consider an STS with $X = \{a\}$ where $a$ is an integer variable that can be non-deterministically incremented by 1, decremented by 1, or held constant at each step. Let the initial state be $a = 1$. Let the property $\Phi$ be $\always (a=1)$, i.e., we require that $a$ is always $1$, which is clearly violated by this system.}

\green{Suppose that $V$ only contains the predicate $lessThanOne[x]$, which returns true if and only if the given integer $x$ is strictly less than $1$. Then, we claim that CCP has no solution.  Indeed, note that the set of counterexample traces includes all traces where at some point either $a<1$ or $a>1$. But the given $V$ is unable to generate an atomic fact where $a$ is greater than $1$ (notice that negation is not allowed in trace constraints). Therefore we cannot classify all counterexample traces, and in particular not those where $a>1$.}

\green{Now suppose that we change $V$ to $\{lessThanOne, greaterThanOne\}$, with the obvious meanings. Then the following two trace constraints constitute a solution to CCP:}
\begin{eqnarray*}
  TC_1 [\rho] & \equiv & \exists i \in [0 .. len(\rho)] : lessThanOne[x@i] \\
  TC_2 [\rho] & \equiv & \exists i \in [0 .. len(\rho)] : greaterThanOne[x@i]
\end{eqnarray*}

\subsection{Uniqueness of Solutions}
\label{subsec-determinism}

\green{The discussion in Section~\ref{subsec-solvability} shows that CCP may or may not have a solution, depending on the set $V$ of predicates allowed in the trace constraints. In this section we show that even for a fixed $V$, CCP does not necessarily have a unique solution.}

\green{Consider the example given just above, in Subsection~\ref{subsec-solvability}. If we set $V$ to $\{lessThanOne$, $greaterThanOne$, $\ne \}$, where $\ne$ is the not-equals predicate, the problem now admits at least two solutions. $W_1$ is still a solution, while the second solution $W_2 = \{TC_3\}$ uses only the $\ne$ predicate to characterize the violating behavior. The trace constraint $TC_3$ is defined as:}
\begin{eqnarray*}
  TC_3 [\rho] & \equiv & \exists i \in [0 .. len(\rho)] : x@i \ne 1 
\end{eqnarray*}

}\fi

\sectionTightenLower

\section{Classification Method}
\sectionTightenLower

In this section, we present a method for solving the CCP introduced in Section~\ref{subsec_ccp}.
We present an overview of our proposed classification algorithm (Section~\ref{sec-alg-overview}), \ifdefined\fulltext{describe }\else{note }\fi optimizations to ensure the generation of a non-redundant classification with minimal classes (Section~\ref{sec-alg-opt}), and finally present a solution to the Running example (Section~\ref{sec-sol}).

\subsectionTightenUpper
\subsection{Algorithm Overview}\label{sec-alg-overview}
\subsectionTightenLower
Given an $\sts$, a property $\Phi$, and a set of user-defined predicates $V$, the goal is to find a set of trace constraints $W$ such that $C(W)$ is a solution to the CCP (Definition~\ref{counterexampleClassificationProblem} in Section~\ref{subsec_ccp}). 
We assume, without loss of generality, that $V$ is non-empty.
Indeed, an empty $V$ implies that the only possible trace constraint is the empty trace constraint, which characterizes the set of all traces. This situation can be modelled by adding to $V$ a trivial predicate that always returns $\True$ (true), thus having a non-empty $V$. 
To guarantee termination, we assume that the set of counterexamples $P = \traces(\sts) \setminus \Phi$ is finite.
To avoid the trivial solution where all traces are violating, we also assume that $\traces(\sts) \cap \Phi\ne\emptyset$.

\begin{figure}[t]
  \begin{algorithm}[H]
    \SetKwProg{Fn}{Func}{}{}
    \DontPrintSemicolon
    \SetKwInOut{Input}{Input}
    \SetKwInOut{Output}{Output}
    \Input{An $\sts$, a specification $\Phi$, and a set of predicates $V$}
    \Output{A set of trace constraints $W$}
    \Fn{$\classify(\sts, \Phi, V)$:}{
      $W = \emptyset$\;
      \While{$\verify(\sts \land \block(W), \Phi)$ == $\nok$}{
        $\rho$ = $\counterexample(\sts \land \block(W), \Phi)$\;
        $\Gamma$ = $\facts(\rho, V)$\;
        \If{$\Gamma = \emptyset$}{
          \KwRet{``$V$ cannot sufficiently characterize the
            violation in $\rho$''}
         }
        $w$ = $\traceConstraint(\Gamma, \rho)$\;
        \If{$\verify(\sts \land w, \neg \Phi) == \nok$}{
          \KwRet{``$V$ cannot sufficiently characterize the
            violation in $\rho$''}
        }
        $w$ = $\minimizetc(\sts, w, \Phi)$\;
        $W$ = $W \cup w$\;
      }
      $W$ = $\redundancyCheck(\sts, W, \Phi)$\;
      \KwRet{$W$}\;
    }
    \caption{The counterexample classification algorithm.}
    \label{alg-classify}
  \end{algorithm}
\end{figure}

The pseudocode for the classification algorithm is shown in Algorithm~\ref{alg-classify}. 
Procedure $\classify$ relies on the existence of a {\em verifier} that is capable of checking $\sts$ against $\Phi$ and generating a counterexample trace, if it exists. In particular, $\classify$ uses the following verifier functions:
\begin{itemize}
\item $\verify(\sts \land \varphi, \Phi)$: Returns $\ok$ if $\sts$ satisfies $\Phi$ under the additional constraint $\varphi$, i.e., if $\traces(\sts\land\varphi) \subseteq \Phi$; else, returns $\nok$. The  constraint $\varphi$ is typically a trace constraint. We  provide examples of $\varphi$ later in this section.
\item $\counterexample(\sts\land\varphi, \Phi)$: If $\verify(\sts\land\varphi, \Phi)$ $==$ $\nok$, returns a trace $\rho$ of $\sts$ such that $\rho \models \varphi$ and $\rho \not\models \Phi$; else, returns an empty output.
\end{itemize}
The algorithm begins by checking whether $\sts$ violates $\Phi$ (line 3) and if so, returning a counterexample that demonstrates how a violation can occur (line 4). The additional argument to the verifier, $\block(W)$, is used to prevent the verifier from re-generating a counterexample that belongs to any previously generated classes; we will describe this in more detail later in this section.

Next, given a particular counterexample $\rho$, the helper function $\facts$ generates the set $\Gamma$ of all atomic facts and atomic position facts that hold over $\rho$, by instantiating the predicates $V$ over the states in $\rho$ (line 5). Then, based on $\Gamma$, $\traceConstraint$ builds a trace constraint that characterizes $\rho$. In particular, this procedure transforms $\Gamma$ into a syntactically valid trace constraint $w$, by (1) introducing a sequence of existential quantifiers over all positional variables in $\rho$ and (2) taking the conjunction of all facts in $\Gamma$ (line 8).

In the next step, the verifier is used once again to ensure that the trace constraint $w$ sufficiently captures the violating behavior in $\rho$ (line 9). This is done by checking that every trace of $\sts$ that satisfies $w$ (i.e., it shares the same characteristics of $\rho$ as described by $\Gamma$) results in a violation of $\Phi$. If not, it implies that $w$ is not strong enough to guarantee a violation; i.e., $V$ does not contain enough predicates to fully characterize $\rho$. In this case, a solution to the CCP cannot be produced and the algorithm terminates with an error (line 10).

If $w$ guarantees a violation,
it is added to the set of classes that will eventually form a solution classification to the CCP (line 12). The process from lines 4 to 12 is then repeated until it exhausts the set of all counterexample classes for $\sts$ and $\Phi$.

To prevent the verifier from returning the same type of counterexample as $\rho$, $\classify$ passes $\block(W)$ as an additional constraint to $\verify$, where:
\begin{align*}
\block(W) \equiv \neg (\bigvee_{i=1}^{|W|} w_i)
\end{align*}
In other words, by including $\block(W)$ as an additional constraint, the verifier ensures that it only explores traces that do not belong to any of the classes in $W$. Note that if $W$ is empty (as in the first iteration of the loop), $\block(W)$ returns \emph{true}  (i.e., $\True$). 

Once the verifier is no longer able to find any counterexample, the algorithm terminates by returning $W$ as the solution classification (line 14).

Provided there is a finite number of counterexamples and a non-empty set of accepting traces, Algorithm~\ref{alg-classify} terminates because at least one counterexample is classified at each iteration of the while loop.
The following theorems establish the correctness of the algorithm.

\begin{theorem}
  Any $W$ returned by $\classify$ is a valid solution to the CCP. 
\end{theorem}

\ifdefined\proofs
        \begin{proof}
We need to show (1) that each $w\in W$ is a trace constraint over the set of state variables $X$ using predicates in $V$ and (2) that $C(W)$ is a classification of the set of counterexamples $P$. (1) follows by construction from Algorithm~\ref{alg-classify}. For (2), note that in order for $C(W)$ to be a valid classification of $P$, it has to cover all the traces in $P$. This follows from the fact that in order for $W$ to be returned, Algorithm~\ref{alg-classify} needs to terminate, which means that the while loop on line 3 exits. This in turn implies that $\verify(\sts \land \block(W), \Phi)$ returns $\ok$, which means $\traces(\sts\land\block(W))\subseteq\Phi$, which implies the result.
\qed
\end{proof}

\fi

\begin{theorem}
If $\classify$ returns no solution (lines 7 or 10 of Algorithm~\ref{alg-classify}), then CCP has no solution for the given $V$.
\end{theorem}

\ifdefined\proofs
        \begin{proof}
We need to show that no solution exists in each of the two cases when $\classify$ returns no solution.

Case 1: $\classify$ returns on line 7. This means that $\Gamma = \emptyset$, i.e., $\facts(\rho, V) = \emptyset$.
From our assumption that there exists an accepting trace, $\rho$ cannot be characterized by the empty trace constraint.
Thus, $\rho$ must be characterized by some non-empty trace constraint $w$. 
Such a $w$ must contain at least one fact that ranges over $X$ (the set of state variables of $\sts$), uses predicates in $V$, and holds over $\rho$. But since $\facts(\rho, V) = \emptyset$, no such facts exist and therefore $w$ cannot exist. 

Case 2: $\classify$ returns on line 10.
This means that $\verify(\sts\land w, \neg\Phi)$ returns $\nok$, i.e., $\traces(\sts\land w) \cap \Phi \ne \emptyset$. 
This in turn means that $w$ does not guarantee a violation of $\Phi$; that is, there are traces of $\sts$ that are characterized by $w$, and yet they satisfy the property $\Phi$. Such traces are therefore not counterexamples.
However, by Definitions~\ref{def:classification} and~\ref{counterexampleClassificationProblem}, each generated trace constraint must characterize a subset of the set of counterexamples, ie. $c(w) \subseteq P$.
If $w$ cannot guarantee a violation this means that $c(w) \not\subseteq P$, so $w$ is not a valid trace constraint.

However, the fact that $w$ is not valid does not immediately imply that there does not exist another trace constraint $w'$ which is valid and characterizes $\rho$. Suppose such a $w'$ exists; that is, suppose that (1) $w'$ characterizes $\rho$ and (2)  $\traces(\sts\land w') \cap \Phi = \emptyset$.
  For $w'$ to characterize $\rho$, $\rho$ must satisfy all the conjuncts of $w'$. 
  Since every conjunct in $w'$ was generated with $V$ and $\rho$ satisfies every conjunct, $w$ should also contain each conjunct in $w'$, by construction of $\facts$ and $\traceConstraint$.
Therefore, the set of conjuncts of $w'$ is a subset of those of $w$, which means that $w'$ is a weaker constraint than $w$. But  
  this contradicts the facts that $\traces(\sts\land w) \cap \Phi \ne \emptyset$ while $\traces(\sts\land w') \cap \Phi = \emptyset$. Indeed, if $w$ allows some traces of $\sts$ which satisfy $\Phi$, and $w'$ is weaker, then $w'$ must also allow those traces, which means that $\traces(\sts\land w') \cap \Phi$ cannot be empty.
  Thus, $\rho$ cannot be characterized by any trace constraint and no solution can be found.
\qed
\end{proof}

\fi

\begin{example} \label{sol-example}
\ifdefined\fulltext
Recall the example from Section \ref{subsec-solvability}.
\else
Consider an STS with $X = \{a\}$ where $a$ is an integer variable that can be non-deterministically incremented by 1, decremented by 1, or held constant at each step. Let the initial state be $a = 1$. Let the property $\Phi$ be $\always (a=1)$, i.e., we require that $a$ is always $1$, which is clearly violated by this system.
\fi
To make $P$ finite, we assume that the length of counterexample traces is exactly $2$. Then, $P = \{(a=1) \lt{--} (a=0), (a=1) \lt{++} (a=2)\}$. Let the set of user-defined predicates be $V$ $=$ $\{lessThanOne, greaterThanOne\}$.

  Suppose that the verifier returns $\rho = (a=1) \lt {--} (a=0)$ as the first counterexample (line 4). Next, $\facts$ evaluates the predicates in $V$ over the state variable $a$ at position $0$ and $1$ (line 5), producing $\Gamma$ that contains one fact: $\{lessThanOne[a@1]\}$. Then, the trace constraint $w$ constructed based on $\Gamma$ is :
  $$TC_1[\rho] \equiv \exists i_1 \in [0 .. len(\rho)]: lessThanOne[a@i_1]$$
It can be shown that any trace of $\sts$ that satisfies $TC_1$ is a violation of $\Phi$; thus, this newly created constraint $w \equiv TC_1$ is added to the set W.  
  



In our example, there is one more counterexample; namely, $\rho = (a=1) \lt{++} (a=2)$, which can be used to construct the following additional trace constraint:
$$TC_2[\rho] = \exists i_1 \in [0 .. len(\rho)]: greaterThanOne[a@i_1]$$
Once $TC_2$ is added to $W$, there are no more remaining counterexamples, and the algorithm terminates by returning $W = \{ TC_1, TC_2\}$.
\end{example}

\vspace{-2mm}
\subsectionTightenUpper
\subsection{Optimizations}\label{sec-alg-opt}
\subsectionTightenLower

\unless\ifdefined\fulltext{Our tool implements multiple optimizations, including trace constraint minimization and redundancy elimination. These topics are treated in the full version of the paper~\cite{arxiv}, which will be available on $\arxivLink$.}\fi

\ifdefined\fulltext{
\subsubsection{Minimizing trace constraints}
\green{A trace constraint $w$ generated on line 6 in Algorithm~\ref{alg-classify} may be a sufficient characterization of $\rho$, but it may also contain facts that are \emph{irrelevant} to the violation. To be more precise, we consider a fact $f \in \Gamma$ to be irrelevant if trace constraint $w$ that is constructed from $\Gamma' \equiv \Gamma - f$ is still sufficient to imply a violation.}

\green{Let us revisit Example~\ref{sol-example}. Suppose that we add to the set $V$ of user-defined predicates an additional predicate $<$ over position variables. Then, for the counterexample $\rho = (a=1) \lt {--} (a=0)$, $\facts$ returns $\Gamma = \{lessThanOne[a@1], 1 < 2\}$ where 1 and 2 are positions in $\rho$. Then, the trace constraint generated by $\traceConstraint$ will be:
$$TC_3[\rho] = \exists i_1, i_2 \in [0 .. len(\rho)]: lessThanOne[a@i_2] \land i_1 < i_2 $$
Although $TC_3$ is sufficient to imply a violation, it
is less general than the previously generated $TC_1$ in the absence of predicate $<$ (see Example~\ref{sol-example}).
Indeed, the constraint $i_1<i_2$ in $TC_3$ forces the condition $a<1$ to occur only at positions $i_2>0$, whereas in $TC_1$ the same condition can also occur at position $i_1=0$. Furthermore, this additional constraint can be safely removed from $TC_3$ while still guaranteeing a violation. Thus, constraint $i_1<i_2$ is an irrelevant fact.}

\green{Our algorithm performs an additional \emph{minimization} step to remove all such irrelevant facts from $w$. This additional procedure provides two benefits: (1) it reduces the amount of information that the user needs to examine to understand the classes and (2) each minimized class is a generalization of the original class and covers an equal or larger set of traces that share the common characteristics, thus also reducing the number of classes in the final classification.}

\green{As shown in Algorithm~\ref{algo_minimizetc}, $\minimizetc$ relies on the ability of certain verifiers (such as the ones based on SAT~\cite{jackson_alloy_2002} or SMT solvers~\cite{de_moura_z3_2008}) to produce a \emph{minimal core} for the unsatisfiability of a formula~\cite{torlak_finding_2008}. In particular, $\minCore(\sts, w, \lnot \Phi)$ computes a minimal subset of conjuncts in the symbolic representation of $\sts$ and $w$ that are sufficient to ensure that $\lnot \Phi$ holds (line 6). The facts ($\gamma$) that are common to this core and $\Gamma$ represent the minimal subset of facts about $\rho$ that are sufficient to imply a violation; a new trace constraint is then constructed based on this subset and returned as the output of $\minimizetc$ (line 7).}

\green{Note that if $\verify$ on line 4 returns $\nok$ (i.e., $\neg \Phi$ does not always hold under constraint $w$), this implies that the set of facts in $\Gamma$ is not sufficient to imply a violation of $\Phi$. However, if $\minimizetc$ is invoked from line 9 in Algorithm~\ref{alg-classify}, this side of the conditional branch should never be reachable.}
 
\begin{figure}[t]
  \begin{algorithm}[H]
    \SetKwInOut{Input}{Input}
    \SetKwInOut{Output}{Output}
    \Input{An $\sts$, a trace constraint $w$, and a specification $\Phi$}
    \Output{A minimized trace constraint}
    \SetKwProg{Fn}{Func}{}{}
    \DontPrintSemicolon
    \Fn{$\minimizetc(\sts, w, \Phi$):}{
      \If{$\verify(\sts \land w), \lnot \Phi)$ == $\ok$}{
        $\gamma$ = $\Gamma \cap \minCore(\sts, w, \lnot \Phi)$\;
        \KwRet{$\traceConstraint(\gamma, \rho)$}\;
      }
      \Else {
       \KwRet{``$\Gamma$ does not sufficiently characterize the
            violation in $\rho$''}\; 
       }
    }
    \caption{$\minimizetc$, which removes from trace constraint $w$
      all facts that are irrelevant to the violation depicted by $\rho$.\label{algo_minimizetc}}
  \end{algorithm}
\end{figure}

\subsubsection{Non-Redundancy}

\green{Although non-redundancy of classification $W$ is not necessary for a valid solution to the CCP, it is a desirable property as it reduces the number of classes that the user needs to inspect. Thus, the main algorithm $\classify$ also performs a redundancy check at its end (line 11, Algorithm~\ref{alg-classify}) to ensure the non-redundancy of any solution that it produces.}

\begin{figure}
    \begin{algorithm}[H]
        \SetKwProg{Fn}{func}{}{}
        \DontPrintSemicolon
        \SetKwInOut{Input}{Input}
        \SetKwInOut{Output}{Output}
        \Input{an $\sts$, a set of trace constraints $W$, and a specification $\Phi$} 
        \Output{a set of trace constraints $W'$}
        \Fn{$\redundancyCheck(\sts, W, \Phi$):}{
          $W'$ = $\emptyset$\;
          \For{$w \in W$}{
            \If{$\verify(\sts \land \block(W \setminus \{w\}), \Phi)$ == $\nok$}{
              $W'$ = $W' \cup w$\;
            }
          }
          \KwRet{$W'$}\;
        }
        \caption{$\redundancyCheck$ checks whether any $w \in W$ is redundant and if it is, removes it.}
        \label{alg-redundancy}
      \end{algorithm}
\end{figure}

  \green{Function $\redundancyCheck$, shown in Algorithm~\ref{alg-redundancy}, ensures that no trace constraint $w \in W$ is covered by any other trace constraints in $W$. 
  Note that when the while loop in Algorithm~\ref{alg-classify} is exited, $\verify(\sts\land \block(W), \Phi)$ returns $\ok$ since $W$ classifies all counterexamples in $P$. This means that all traces of $\sts$ which do not belong in any of the classes in $W$ satisfy $\Phi$. To find redundant trace constraints, we iterate over each $w \in W$ and check whether $\sts$ still satisfies $\Phi$ with $w$ removed from $W$ (line 4, Algorithm~\ref{alg-redundancy}). If this is the case, then $w$ is redundant, since $W\setminus\{w\}$ already covers $P$. Otherwise, $w$ must characterize some $\rho \in P$ that the other trace constraints do not, and thus $w$ is added to the non-redundant set $W'$, which is returned at the end.}

\green{For example, recall the predicates $V = \{$$\ne$, $lessThanOne$, $greaterThanOne$$\}$ from Section~\ref{subsec-determinism}. Suppose that $\classify$ finds two trace constraints in this order\footnote{Note that a newly created trace constraint is never redundant.}:}
\begin{eqnarray*}  TC_1
  [\rho] & \equiv & \exists i \in [0 .. len(\rho)] : lessThanOne[a@i]\\
  TC_2
  [\rho] & \equiv & \exists i \in [0 .. len(\rho)] : a@i \ne 1
\end{eqnarray*}
\green{Notice that $TC_2$ classifies all counterexamples that $TC_1$ classifies.
Thus, $TC_1$ is redundant and is not added to the final solution $W'=\{TC_2\}$.}

}\fi

\subsectionTightenUpper
\subsection{Solution to the Running Example}\label{sec-sol}
\subsectionTightenLower
Consider the running example presented in Section~\ref{sec_running_example}.
For this example, Algorithm~\ref{alg-classify} outputs the trace constraints $TC_{Encrypted}$ and $TC_{Plaintext}$ in Section~\ref{sec_running_example} given
the set of predicates $V = \{=, <\}$. Equality $=$ operates over $\Message$s and Booleans while $<$ operates on position variables.

Atomic position facts are generated just like atomic facts. Recall the following counterexample trace that is characterized by $TC_{Encrypted}$: 
$$\rho = (\emptyset,\False) \lt{(\Encrypted,\Alice,\False)} (\Key,\False) \lt{(\Encrypted,\Alice,\True)} (\Key,\True)$$

In the $\facts$ procedure, the $<$ predicate would generate two facts, $\{i_1<i_2, i_2<i_3\}$. These facts impose an ordering on any satisfying counterexample and capture the timing of the violation.

\sectionTightenLower

\section{Implementation and Case Studies}
\label{sec_case_study}
\sectionTightenLower
\subsection{Implementation}
\subsectionTightenLower

We have built a prototype implementation of the $\classify$ algorithm (Algorithm~\ref{alg-classify}) on top of the Alloy Analyzer\cite{jackson_alloy_2002}, a formal modeling and verification tool.  In particular, Alloy uses an off-the-shelf SAT solver to perform bounded model checking (BMC), which is used for the $\verify$ procedure in the algorithm.  As we demonstrate in this section, our prototype is capable of characterizing a large set of counterexamples (hundreds of thousands) with only a handful of generated classes.  These generated classes are provided to the user in the form of trace constraints, along with representative counterexamples from each class.

Even though our current implementation uses Alloy and BMC, our technique does not depend on the use of BMC or any particular verification engine and could be implemented using other tools, provided they are capable of generating counterexample traces. Our current implementation does rely on the SAT solver being able to compute minimal unsatisfiable cores (which are used for minimizing the trace constraints\ifdefined\fulltext{).}\else{, as explained in more detail in the full version~\cite{arxiv}).}\fi

\subsectionTightenUpper
\subsection{Case Studies: Needham-Schroeder}
\subsectionTightenLower

As a case study, we applied our prototype to the well-known Needham-Schroeder protocol (NSP)~\cite{needham_using_1978}, which has been known to be vulnerable to certain types of attacks~\cite{lowe_attack_1995}. We show how our classification methods can be used to classify the large number of counterexamples in a formal model of NSP into a small number of classes that correspond to these types of attacks.

The purpose of NSP is to allow two parties to communicate privately over an insecure network. NSP has two variants that look to accomplish this goal in different ways. The first variant is the Needham-Schroeder Symmetric protocol, from now on referred to as Symmetric, and the second variant is the Needham-Schroeder Public-Key protocol, from now on referred to as Public-Key. The two variants exhibit different violating behaviors, which allowed us to test our classification technique on the two separate variants, while not having to write two drastically different models.

\unless\ifdefined\fullText{\vspace{-4mm}}\fi
\subsubsection{Formal Modeling}
\subsectionTightenLower

We constructed Alloy models of both the Symmetric and Public-Key variants. Together, both variants total approximately 700 lines of Alloy code.
These models serve as the input to our tool along with a specification $\Phi$ and a set of predicates $V$\footnote{The Alloy models and code for our tool can be found at \url{https://github.com/cvick32/CounterexampleClassificiation}.}.

In both variants there are 4 \Process{es}:  \Alice, \Bob, \Eve, and a central \Server. The attacker, \Eve{,} can read all of the \Message{s} exchanged between the \Process{es}. The setup is  similar to the running example that has been discussed throughout the paper. Both variants must  satisfy the following specification.

\para{Specification ($\Phi$)}

We consider only one property across both variants of NSP: 
the secret $\texttt{Key}$ $K_{AB}$ shared between \Alice and \Bob is not leaked to \Eve.
 We express this property as the following LTL formula:
$$ \Phi = \always(K_{AB} \not\in Eve.knows)$$		
where $p.knows$ denotes the state variable of a protocol participant representing the set of $\texttt{Key}$s that the participant $p$ has access to.

\para{Symmetric}

In the Symmetric variant, \ifdefined\fulltext{illustrated in Fig.~\ref{fig:needhamSchroeder},}\fi \Alice notifies the \Server that she would like to communicate with \Bob. The \Server then generates a communication key, \Key{,} for \Alice and \Bob and sends it to \Alice{.} 
This message is encrypted with \Bob{'}s secret key. \Alice forwards this message to \Bob so that he will be able to decrypt the message with his secret key and learn the shared key. \Bob then sends a random nonce to \Alice that is encrypted with their shared key. \Alice verifies that she knows the shared key by sending back \Bob{'}s nonce decremented by 1. 

\ifdefined\fulltext{
\begin{figure}
  \centering
  \begin{adjustbox}{max totalsize={1.0\textwidth}{.5\textheight}}
    
    \begin{tikzpicture}[
        every node/.append style={very thick,rounded corners=0.1mm}]

        \node[draw,rectangle] (Alice) at (0,0) {Alice};
        \node[draw,rectangle] (Server) at (5.5,0) {Server};
        \node[draw, rectangle] (Bob) at (7.5,0) {Bob}; 
        \node[draw, rectangle] (Eve) at (9,0) {Eve};


        \draw [thick] (Alice)--++(0,-6);
        \draw [thick] (Server)--++(0,-2);
        \draw [thick] (Bob)--++(0,-6);
        \draw [thick] (Eve)--++(0,-6);
        
        \draw [->,very thick] (0,-1)--node [auto] {$\{A,B,N_A\}$}++(5.5,0);
        \draw [<-,very thick] (0,-2)--node [auto] {$\{N_A, K_{AB}, B, \{K_{AB}, A\}_{K_{BS}}\}_{K_{AB}}\}$}++(5.5,0);

        \draw [->, very thick] (0,-3.5)--node [auto] {$\{K_{AB}, A\}_{K_{BS}}$}++(7.5,0);
        \draw [<-, very thick] (0,-4.5)--node [auto] {$\{N_B\}_{K_{AB}}$}++(7.5,0);
        \draw [->, very thick] (0,-5.5)--node [auto] {$\{N_B - 1\}_{K_{AB}}$}++(7.5,0);

        \draw[->,snake=snake,very thick,draw=red,segment amplitude=.4mm,segment length=2mm,line after snake=1mm] (5.5,-1)--node [auto] {}++(3.5,0);
        \draw[->,snake=snake,very thick,draw=red,segment amplitude=.4mm,segment length=2mm,line after snake=1mm] (5.5,-2)--node [auto] {}++(3.5,0);
        \draw[->,snake=snake,very thick,draw=red,segment amplitude=.4mm,segment length=2mm,line after snake=1mm] (7.5,-3.5)--node [auto] {}++(1.5,0);
        \draw[->,snake=snake,very thick,draw=red,segment amplitude=.4mm,segment length=2mm,line after snake=1mm] (7.5,-4.5)--node [auto] {}++(1.5,0);
        \draw[->,snake=snake,very thick,draw=red,segment amplitude=.4mm,segment length=2mm,line after snake=1mm] (7.5,-5.5)--node [auto] {}++(1.5,0);

    \end{tikzpicture}
  \end{adjustbox}
  \caption{A communication diagram of the Needham-Schroeder Symmetric protocol. $A$ and $B$ are identifiers for $\Alice$ and $\Bob$ respectively. There are three keys: $K_{AB}$, the shared key between $\Alice$ and $\Bob$, $K_{AS}$ and $K_{BS}$ which are each $\Alice$ and $\Bob{'s}$ server key. $\Alice$ and $\Bob$ also make use of a nonce, $N_A$ and $N_B$ respectively. Each arrow reprepresents a $\Message$. $\{...\}_{K}$ denotes a $\Message$ encrypted by key $K$, and therefore requiring $K$ to be read successfully. The snaking red lines represent $\Eve$ having access to all $\Message{s}$ that are sent over the network.}
  \label{fig:needhamSchroeder}
\end{figure}
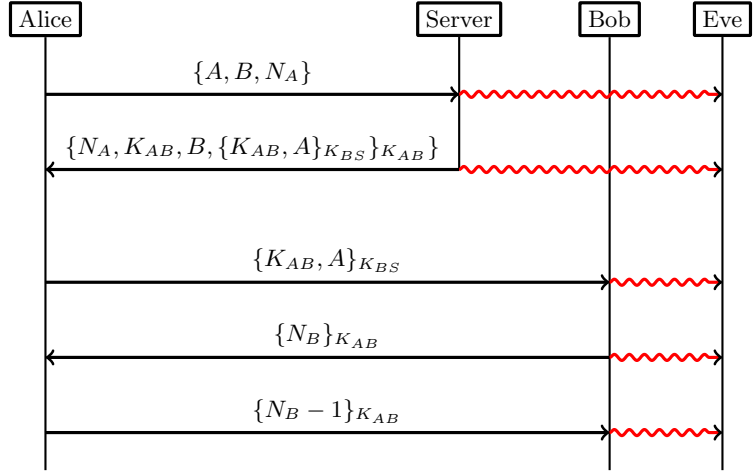

}\fi

\para{Public-Key}

In the Public-Key variant, \Alice notifies the \Server that she would like to communicate with \Bob. The \Server sends \Alice a signed message with \Bob{'s} public key. \Alice sends \Bob a message including a nonce that is encrypted with \Bob{'s} public key. \Bob receives this message and asks the \Server for \Alice{'s} public key. The \Server sends \Bob \Alice{'s} public key. \Bob now sends \Alice{'s} nonce back to \Alice along with a new nonce encrypted with \Alice{'s} public key. \Alice confirms that she has her private key by responding to \Bob with his nonce encrypted with his public key. 

\para{Predicates}

In the experiments described below, we used the following sets of predicates ($V$):
$\Generic = \{=, <\}$, consisting of only equality
and one ordering predicate;
$V_1=\Generic \cup \{\replay\}$;
and
$V_2=\Generic\cup \{\manInTheMiddle\}$.
$V_1$ and $V_2$  include all generic predicates plus some specialized predicates that characterize particular behavior in a model.
\ifdefined\fulltext
The $\replay$ predicate, shown in Fig.~\ref{fig:replay}, 
\else
The $\replay$ predicate 
\fi
captures counterexamples where $\Eve$ sends the same message that was sent earlier by another process.
The $\manInTheMiddle$ predicate captures counterexamples where $\Eve$ passes $\Alice$ and $\Bob$'s messages between them with no direct communication between $\Alice$ and $\Bob$. 

Predicates like $\replay$ and $\manInTheMiddle$ could be part of a library of predicates that any user could search and use. For example, $\replay$ can be used to check other communication protocols for replay attacks, provided that they follow a similar message-passing structure. Note that no information concerning the particularities of the Needham-Schroeder protocol is used in the definition of $\replay$, meaning that this predicate can be used in a generic way. The same holds for $\manInTheMiddle$.


\ifdefined\fulltext{
\begin{figure}[t]
  \begin{algorithm}[H]
    \SetKwProg{Fn}{pred}{}{}
    \DontPrintSemicolon
    \SetKwInOut{Input}{Input}
    \SetKwInOut{Output}{Output}
    \Input{A counterexample $\rho$ and two time indexes $t1$ and $t2$}
    \Output{A boolean}
    \Fn{$\replay[\rho, t1, t2]$:}{
      $t1 < t2 \, \land$\;
      $\rho.msg.sender.t1 \ne \Eve \xspace \, \land$\;
      $\rho.msg.sender.t2 = \Eve \, \land$\;
      $\rho.msg.nonce.t2 = \rho.msg.nonce.t1 \, \land$\;
      $\rho.msg.process.t2 = \rho.msg.process.t1 \, \land$\; 
      $\rho.msg.key.t2 = \rho.msg.key.t1 \, \land$\;
      $\rho.msg.encryption.t2 = \rho.msg.encryption.t1 \, \land$\;
    }
  \end{algorithm}
  \caption{The $\replay$ predicate returns $\True$ if there are two positions $t1$ and $t2$ in $\rho$ such that $t1$ occurs before $t2$ and the $\Message$ at $t1$ is the exact same as the $\Message$ at $t2$ except that $\Eve$ is now the sender.}
  \label{fig:replay}
\end{figure}

}\fi


\unless\ifdefined\fulltext{\vspace{-4mm}\fi
\subsubsection{Results}
\subsectionTightenLower
Our tool was able to produce classifications for both the Symmetric and Public-Key variants of NSP, as explained below. We were able to count up to $270,000$ counterexamples (using the counterexample enumeration feature in Alloy) for both NSP variants until our program ran out of memory. The results are shown in Table~\ref{pubTable}. 

Alloy employs bounded model checking for its verification engine; the {\em bound} column in Table~\ref{pubTable} shows the upper bound used for the number of steps in traces explored by BMC.
The $V$ column shows the predicate set used in each experiment.
The next column shows the number of classes generated and the last two columns show the execution time in seconds\footnote{Times were measured using the Java built-in \texttt{System.nanoTime()}.}.
The execution time is split into the time our tool spent calling Alloy to find counterexamples and all other computations on the right.
We found it instructive to show that the program was spending much of its time generating counterexamples in Alloy, while all other computations remained relatively constant for each respective experiment.
Note that executions using $V_2$ take much longer than other executions. Most of this time is spent in generating the facts for $\manInTheMiddle$ as that particular predicate ranges over a number of time steps and all time steps in a counterexample must be checked.
We also note that when using the $\Generic$ predicate set no redundant classes were found.

\begin{table}[h]
  \centering
  \resizebox{2.25in}{!}{
  \begin{tabular}{|c|c|c|c|c|}
    \hline
    bound & $V$ & \# classes & Alloy time & Total time \\
    \hline
    \multirow{2}{1em}{10} & $\Generic$ & 2 & 1.92 & 7.56 \\ 
      & $V_1$     & 3 &            4.29 & 10.23 \\ \hline
    \multirow{2}{1em}{25} & $\Generic$ & 2 & 9.37 & 16.24 \\ 
      & $V_1$     & 3 &            37.26 & 43.55 \\ \hline
    \multirow{2}{1em}{50} & $\Generic$ & 2 & 61.48 & 70.41 \\
    & $V_1$     & 3 &           220.49 & 226.37 \\ \hline
    \multirow{2}{1em}{75} & $\Generic$ & 2 & 254.38 & 267.96 \\
       & $V_1$     & 3 &           897.19 & 903.44 \\ \hline
    \multirow{2}{1em}{100} & $\Generic$ & 2 & 653.62 & 674.66 \\
       & $V_1$     & 3 &            1949.65 & 1955.133 \\ \hline 
  \end{tabular}
  }
  \label{symTable}
  \quad
  \resizebox{2.25in}{!}{
  \begin{tabular}{|c|c|c|c|c|}
    \hline
    bound & $V$ & \# classes  & Alloy time & Total time \\
    \hline
    \multirow{2}{1em}{10} & $\Generic$ & 2 & 2.96 & 12.46 \\ 
    & $V_2$ & 3 &  6.01 & 100.61 \\ 
    \hline
    \multirow{2}{1em}{25} & $\Generic$ & 2 & 12.96 & 24.04 \\
    & $V_2$ & 3 & 30.64 & 125.70 \\
    \hline
    \multirow{2}{1em}{50} & $\Generic$ & 2 & 91.67 & 97.78 \\
    & $V_2$ & 3 & 157.62 & 251.89 \\
    \hline
    \multirow{2}{1em}{75} & $\Generic$ & 2 & 321.95 & 349.21 \\
    & $V_2$ & 3 & 525.53 & 615.85 \\
    \hline
    \multirow{2}{1em}{100} & $\Generic$ & 2 & 850.52 & 893.71 \\
    & $V_2$ & 3 & 1301.83 & 1396.92 \\ \hline
  \end{tabular}
  }
  \caption{Results on the Symmetric (left) and Public-Key (right) NSP variants. All times are recorded in seconds. All experiments were evaluated on a a 2.5GHz Quad-Core Intel i7 CPU with 16GB of RAM.}
  \label{pubTable}
\end{table}

\para{Symmetric}

This NSP variant is vulnerable to a replay attack. This attack has been addressed in implementations like Kerberos, although the attack was not found until 3 years after the initial publication of the protocol \cite{denning_timestamps_1981}. 

Using the $\Generic$ predicate set, our tool generated 2 non-redundant classes. 
These classes characterize counterexamples where either $\Alice$ or $\Bob$ unknowingly establishes communication with $\Eve$, who then manages to extract the secret key from this interaction.  For example, the trace constraint $TC_{Generic}$ shown below represents one of these two classes and characterizes counterexamples where
$\Alice$ sends a message and at a later state, $\Eve$ manages to learn the secret key:
\begin{eqnarray*}
  TC_{Generic}
  [\rho] \equiv \exists i_1, i_2 \in [0 .. len(\rho)] : & & \rho.msg.sender@i_1  = \Alice \land \\
  & & \rho.\Eve.knows@i_2 = \{ \Key \} \land i_1 < i_2 
\end{eqnarray*}
Although this constraint is a valid characterization of counterexamples (in that it is sufficient to guarantee a violation of $\Phi$), it is rather an abstract one, in that it does not describe the intermediate steps that $\Eve$ carries out in order to extract the secret key.

To generate more specialized classes, the user can provide additional predicates beside the generic ones.  Using $V_1$ as the predicate set, our tool generated 3 classes: the two classes previously found with $\Generic$, plus a third class 
represented by the trace constraint $TC_{Replay}$ shown below:
\begin{eqnarray*}
  TC_{Replay}
  [\rho] \equiv \exists i_1, i_2 \in [0 .. len(\rho)] : & & \replay[\rho, i_1, i_2] \land i_1 < i_2  \land \\
  & & \rho.msg.encryption@i_2  = \rho.msg.encryption@i_1 \land \\
  & & \rho.msg.key@i_2 = \rho.msg.key@i_1 
\end{eqnarray*}
\ifdefined\fulltext{
  \green{Our tool guarantees that we begin our classification with counterexamples that satisfy whichever predicate we choose, in this case $\replay$. This is helpful as it constrains our classification to only those counterexamples which satisfy $\replay$, allowing us to classify a subset of the total set of counterexamples.}
}\fi
The constraint $TC_{Replay}$ describes the type of violation where $\Eve$ carries out a replay attack, where she re-sends the message that was previously sent at step $i_1$ again at step $i_2$ with the identical message content. Note that although $TC_{Replay}$ is a redundant class with respect to the other two classes generated using the generic predicates, it serves additional utility in that it provides more specific information about what $\Eve$ does in order to cause a security violation. The user of our tool (e.g., a protocol designer) could then use the information in these constraints to improve the protocol and prevent these types of violations.

\para{Public-Key}

This NSP variant is vulnerable to a man-in-the-middle attack \cite{lowe_attack_1995}. \Eve is able to forward messages between \Alice and \Bob and trick them into thinking they are communicating directly.

Similarly to the Symmetric variant, we were able to classify counterexamples that demonstrated the man-in-the-middle attack. The classes found in the Public-Key experiment reflected what we found in the Symmetric variant, i.e. 2 classes that show a general violating pattern with $\Generic$ and then 3 classes where 1 class demonstrates the known violation, using predicate set $V_2$. Our tool showed that the Public-Key variant is not vulnerable to replay attacks.


In summary, our classification method (1) significantly reduces the amount of information that the user needs to inspect to understand the different types of violations, by collapsing the large number of counterexamples ($\geq$ 270,000 for the case study) into a small number of classes and (2) enables the user to inspect these different violating behaviors in a high-level representation (i.e., trace constraints) that can encode domain-specific information (e.g., replay attacks).


\sectionTightenLower

\section{Related Work}
\sectionTightenLower

It is well known that predicates can be used to abstract needless detail in certain problem domains~\cite{cousot_abstract_1977}~\cite{jhala_predicate_2018}. This is the first time, to our knowledge, that predicates have been used for counterexample classification.

Our work can be considered a kind of automated debugging technique \cite{debuggingbook2021} in the context of model checking.
There have been a number of prior works into locating the relevant parts of counterexample that explain or even \emph{cause} a violation \cite{ball_symptom_2003, groce_what_2003, beer_explaining_2009}. While our work does not deal with an explicit notion of causality, the generated trace constraints are sufficient to imply a violation of the property. The major difference between these works and ours is that they focus on \emph{explaining} one or more given counterexamples, while our objective to \emph{classify} the set of all counterexamples into distinct classes. 
Our work is also related and complementary to \cite{kashyap_producing_2008}, which focuses on generating short counterexamples. 
We take a different approach by focusing on generating minimal trace constraints, each of which characterize a {\em set} of counterexamples.

The approach in \cite{dominguez_generating_2013} has the similar goal of generating a {\em diverse} set of counterexamples. This work relies on a notion of diversity that depends on general properties about the structure of the given state machine (e.g., counterexamples that have different initial distinct and final states). In comparison, our notion of diversity is \emph{domain-specific}, in that it is capable of classifying traces based on domain-specific predicates that can be provided by the user. In this sense, these are two complementary approaches and could potentially be combined into a single model debugging tool.

\sectionTightenLower

\section{Conclusion and Future Work}
In this paper, we have proposed \emph{counterexample classification} as a novel approach for debugging counterexamples generated by a model checker. The key idea behind our approach is to classify the set of all counterexamples to a given model and a property into \emph{trace constraints}, each of which describes a particular type of violation. Our work leverages the notion of \emph{predicates} to distinguish between different types of violations; we have also demonstrated how these predicates can capture violations that are common within a domain (e.g., attacks on security protocols)  and can facilitate the reuse of domain knowledge for debugging.

For future work, we plan to explore methods based on machine learning (such as clustering (e.g., \cite{song_trace_2009}) to automatically extract predicates from a given set of counterexample traces. Another interesting direction is to explore how our classification method could be used to improve counterexample-guided approaches to program synthesis  (such as CEGIS~\cite{solar-lezama_combinatorial}), by reducing the number of counterexamples that need to be explored by the synthesis engine. 


\sectionTightenLower


\bibliography{counterexampleClassification}{}
\bibliographystyle{plain}
\end{document}